\documentclass[a4paper,fleqn,usenatbib]{mnras}

\usepackage{newtxtext,newtxmath}

\usepackage[T1]{fontenc}
\usepackage{ae,aecompl}

\usepackage{graphicx, epstopdf}	
\usepackage{amsmath}	
\usepackage{amssymb}	
\usepackage{xcolor}
\usepackage[normalem]{ulem} 

\newcommand{\art} {{\small ART}}

\newcommand{\hii} {H{\sc ii}}
\newcommand{\kms}{${\rm km~s}^{-1}$}

\newcommand{\Msun}{$M_{\odot}$}

\newcommand{\ndens} {{n_{\rm dens}}}
\newcommand{\nsf} {$n_{\rm SF}$}
\newcommand{\pcc} {{\rm cm}^{-3}}

\newcommand{\sfrt} {${\rm SFR}_t$}

\newcommand{\tage} {t_{\rm age}}
\newcommand{\tff} {t_{\rm ff}}

\title[Shaping of stellar clusters by feedback] {The effect of
photoionising feedback on the shaping of hierarchically-forming stellar clusters}

\author[Gonz\'{a}lez-Samaniego, A., et al.]{
Alejandro Gonz\'{a}lez-Samaniego\thanks{E-mail: a.gonzalez@irya.unam.mx}
Enrique Vazquez-Semadeni 
\\
Instituto de Radioastronom\'ia y Astrof\'isica, Universidad
  Nacional Aut\'onoma de M\'exico, Apdo. Postal 3-72, Morelia, 58089, M\'exico \\
}

\date{\today}

\pubyear{2020}

\begin{document}
\label{firstpage}
\pagerange{\pageref{firstpage}--\pageref{lastpage}}
\maketitle

\setlength\topmargin{-2pc}
\volume{{\rm in press}}
\begin{abstract}

  We use two hydrodynamical simulations (with and without
    photoionising feedback) of the self-consistent evolution of
  molecular clouds (MCs) undergoing global hierarchical collapse
  (GHC), to study the
  effect of the feedback on the structural and
  kinematic properties of the gas and the stellar clusters formed
  in the clouds. During this early stage, the
  evolution of the two simulations is very similar (implying that the
  feedback from low mass stars does not affect the cloud-scale
  evolution significantly) and {\it the star-forming region accretes
    faster than it can convert gas to stars,
      causing the instantaneous measured star formation efficiency
    (SFE) to remain low} even in the absence of significant
  feedback.  Afterwards, the ionising feedback first destroys the
  filamentary supply to star-forming hubs and ultimately removes the
  gas from it, thus first reducing the star formation (SF) and finally halting it.
  The ionising feedback also affects the
  initial kinematics and spatial distribution of the forming stars,
  because the gas being dispersed continues to form stars, 
  which inherit its motion. In the non-feedback
  simulation, the groups remain highly compact and do not mix, 
  while in the run with feedback, the
    gas dispersal causes each group to expand,
  and the cluster expansion thus consists of the combined expansion of
  the groups. Most secondary star-forming sites around
  the main hub are also present in the non-feedback run, implying a
  primordial rather than triggered nature. We do
  find one example of a peripheral star-forming site that appears only
  in the feedback run, thus having a triggered origin. However, this appears to be the
  exception rather than the rule, although this may be an
  artifact of our simplified radiative transfer scheme.  
\end{abstract}

\begin{keywords}
gravitation -- hydrodynamics -- stars: formation -- galaxies: star clusters: general --  ISM: clouds
\end{keywords}

\section{Introduction}
\label{intro}

Stellar groups are the result of star formation (SF) happening within
dense molecular gas.  However, it remains a challenge to have a
complete description for the origin and the structural properties of
stellar clusters (SCs), such as mass segregation, density profiles, age
gradients and histograms, hierarchical structure, etc. 

From the observational point of view, great advances have been
achieved in the study of kinematics of SCs with data from surveys like
GAIA-DR2 \citep{Kounkel+18} and APOGEE-2 \citep[see e.g.,][]{Hacar+16,
DaRio+17, Kounkel+18}. In particular, \citet{DaRio+17} found evidence
for kinematic subclustering and by studying the kinematic properties
through radial velocities they found also some evidence for ongoing
expansion; \citet{Hacar+16} have suggested that for most of the Orion
A cloud, young stars keep memory of the parental gas substructure
where they originated, and \citet{Kounkel+18} have identified distinct
groups of stellar objects with ages ranging from 1 to 12 Myr within
the Orion Complex. These authors also found subclusters and reported
that, while in Orion D and $\lambda$ Ori the motions of the stars are
consistent with an expansion process, Orion B is still in the process
of contraction. Finally, \citet{Kounkel+18} also suggest that the
proper motions in $\lambda$ Ori are consistent with a radial expansion
due to a supernova explosion. These features require a coherent
explanation within current models of formation of SCs and the MCs
where they were formed. In particular, it is important to determine
the role that stellar feedback from massive stars plays on this.

On the theory side, a variety of numerical models of star formation in
the interstellar medium has been used to explain the properties of
SCs following the complete evolution of molecular clouds since their
own formation to the formation of a stellar group within the
region. In particular, the effect of different sources of feedback and
the positioning of the sources has been a matter of intense
research. \citep[see e.g.,][]{Bate09, VS11, Colin+13, Dale+13b,
Dale+14, Iffrig+15, Iffrig+17, Kortgen+16, VS17, Colling+18, Haid+18,
Grudic+18, ZA+19b}. In particular, \citet{Haid+18} showed that the
relative impact of stellar winds and ionising radiation depends on the
stellar mass considered but even more strongly on the properties of
the ambient medium. They found that for stars placed in the CNM, ionising
radiation is the relevant source of feedback to take into account.

One important aspect, which perhaps has not received sufficient
attention, is the need to investigate clouds in which both the
geometry of the clouds and the positioning of the ionising feedback
sources arise self-consistently, in order to avoid unrealistic
setups. Numerical simulations of the formation and evolution of the
clouds in the presence of self-gravity indicate that the clouds engage
in global hierarchical collapse \citep[GHC;] [and references therein]
{VS+19}, forming filamentary accretion flows that transfer mass from
the large to the small scales \citep[e.g.,] [] {Heitsch+08,
Heitsch+09, Smith+09, GV14}. For example, clouds formed self-consistently by
collision of diffuse gas streams tend to be flattened and porous, with
warm, diffuse gas ``pockets'', rather than roundish and isothermal,
and to develop turbulence due to the combined effect of various
instabilities \citep[e.g.,] [] {Walder+98, BP+99, Hennebelle+00,
Hennebelle+08, Audit+05, Heitsch+05, Heitsch+06, VS+06}. A realistic
shape and porous constitution may be essential for the correct
simulation and dispersal of the clouds, as discussed by
\citet{Colin+13}, \citet{Dale15}, and \citet{Raposo+15}.

In this paper we study structural properties of stellar clusters and
the gas around them using hydrodynamical simulations in which the
clouds have formed and evolved self-consistently from the collision of
oppositely-directed streams of diffuse warm gas, with an additional
moderate turbulent component to break the symmetry of the stream
collision setup. The clouds then follow an evolutionary path
consisting of growth, onset of global hierarchical collapse (GHC),
accelerating star formation, cluster formation and cloud dispersal, as
described in \citet{VS+19}.

In these simulations, we test the role of stellar
feedback in shaping the properties of resulting stellar clusters and
the gas that led to its formation and evolution. For this, we use
hydrodynamical simulations presented first in 
\citet[] [hereafter Paper I] {Colin+13} 
and then studied in the context of cluster formation in
\citet[] [hereafter Paper II] {VS17}. 
We use the LAF1 simulation described in Paper I (for ``large-amplitude
fluctuations, feedback on''), focusing, in particular, on the cluster
labeled G1-2 in Paper II. In the present paper, we investigate the
effects of feedback in detail, by including the non- feedback
counterpart of the LAF1 simulation, denoted LAF0.
Besides pinpointing the precise effect
of the feedback on the cluster structure, our results can additionally
help to test the GHC scenario
of MCs by testing its ability to form realistic
clusters, and predict their observed properties.

This article is organized as follows.  In Sec.\ \ref{sims} we present
our simulations and describe the main physical processes included in
the numerical code, and in Sec.\ \ref{proc} we discuss the procedure
to analyze cluster properties. Next, in Sec.\ \ref{results} we present
our main results, which include an analysis of the feedback effect on
gas, stars, and the SFR in the SC. In Sec.\ \ref{disc} we discuss the 
implications of our results and their comparison with other studies. 
Finally, in Sec.\ \ref{conclusions} we summarize our findings and present
our conclusions.

\section{Numerical setup and method}
\label{method}

In this section we first describe the numerical simulations, albeit
only briefly, as they have been already described in detail in Paper I
and Paper II, to which we refer the reader for further details. We
then describe the cluster-identification algorithm.
 
\subsection{Simulations}
\label{sims}
We use the Hydrodynamics+N--body Adaptive Refinement Tree (\art ) code
\citep*{Kravtsov+97, Kravtsov+03} to run the simulations. The initial
conditions consist of two oppositely directed cylindrical
streams that collide within a 256
pc numerical box containing a total mass of $9.25\times
  10^{5}$ \Msun. The gas properties of the uniform background
resemble the conditions of the warm neutral medium (WNM): it has a
uniform density $n = 1\ \pcc$ and temperature $T = 5000 $ K. The
streams, which have the same density as the background,
  have a length of 112 pc, a radius of 32 pc, and travel at a speed
of 5.9 \kms\ each. This velocity is subsonic (Mach number of 0.8)
with respect to the adiabatic sound speed of the WNM. The head-on
collision of the streams promotes a transition to the cold phase,
forming an initially circular sheet of cold, dense gas (cold
  neutral medium), whose thickness and mass grow as the warm gas from
  the streams accretes onto the layer and condenses to the cold phase
  \citep{VS+06}. A turbulent initial velocity field with rms velocity
dispersion of 1.7 \kms\ is added in order to break the symmetry of the
colliding streams and trigger various instabilities within the
  layer \citep{Heitsch+06}, which cause it to bend and fragment. This setup is
  meant to represent the formation of a large cloud by general
  transonic or supersonic compressions in the WNM of either turbulent
  or gravitational origin \citep[e.g.,] [] {vonWeizsacker51,
    Roberts69, Sasao73, Elmegreen+77, Vishniac94, Ballesteros+99a,
    Hennebelle+00, KI02, Audit+05, Heitsch+05, Heitsch+06, VS+06, VS+07}.

As the simulation evolves, and once the mass is redistributed forming
cold dense regions, the simulation allows up to five refinement
levels, reaching a minimum cell size of 0.0625 pc ($\approx$13 000
au). Cells in the mesh are refined when the gas mass within the cell
is greater than 0.32 \Msun. It is important to note that the cell's
mass can reach much larger values than this `refining mass' after the
maximum refining level is reached due to our probabilistic SF scheme
(see \ref{Sform} below).

As described in Paper I, we modified our original version of \art\
such that our simulations include: i) a new probabilistic SF
prescription; ii) self-gravity from gas and stars; iii) parametrized
heating and cooling, and iv) an ionisation feedback prescription by
massive stars. We use the cooling and heating functions provided in
\citet{KI02},

\begin{align}
\frac{\Lambda(T)}{\Gamma}  & =   10^{7} \exp \left( \frac{-1.184 \times 10^{5}}{T + 1000}\right)  \nonumber  \\
              &        \;\;\;\;    +1.4 \times 10^{-2}\sqrt{T}\exp\left( \frac{-92}{T}\right) \rm{cm}^{3} 
\end{align}
\begin{equation}
\Gamma = 2.0 \times 10^{-26} \rm{erg s}^{-1}. 
\end{equation}

This particular parametrization differs from the one presented in
\citet{KI02} since we have included the corrections to the
typographical errors that appeared in the original version, as
discussed in \citet{VS+07}. Under the
conditions imposed by these functions the gas is thermally unstable in
the density range 1 $\lesssim n \lesssim $ 10 $\rm{cm}^{3}$.

\subsubsection{Star formation: a probabilistic approach}
\label{Sform}

For each timestep of the coarsest grid, when the gas density $n$ in a
grid cell exceeds a density threshold \nsf, a stellar particle (SP)
with mass $m_{\it SP}$ may be placed in the cell with a probability
$P$. Indeed, depending on the value of $P$, there is a non-zero
probability (1-$P$) of not forming an SP, even when the density cell
reaches \nsf. If this is the case, then the gas accretion into the
cell continues for several timesteps until a SP is created. When an SP
finally forms, it acquires half the gas mass of its parent cell, and
does not accrete further. 
Thus, a cell that exceeds the density threshold but does not
 form a star can increase its density and mass until it eventually 
 does form a star. This mimics the process of
 accretion onto a protostar to 
 form a massive star. Indeed, the prescription we use implies that 
 the longer it takes to form an SP in a collapsing cell, the more 
 massive the SP will be, because the density of the cell will be higher.

Consequently, our probabilistic prescription allows the formation of SPs with
different masses. In Paper I it was shown that the resulting mass
distribution of the SPs is a power law with an exponent than can be
tuned depending on the value of $P$.  In particular, it was found in
Paper I that, for $P = 0.003$, the slope of the SP mass distribution
(the simulation's IMF) in run LAF1 simulation went from $-1.21$ at
$\approx$ 6 Myr after the SF starts,  to $-1.34$ at the end of the
evolution, $\approx$ 21 Myr after the first star was
formed, thus resembling the Salpeter IMF. Also, for our fiducial
setup, with five refinement levels, $P = 0.003$, and \nsf $= 9.2 \times
10^{4}\; \pcc$, the minimum SP mass in the simulation was $m_{\rm SP} =
0.39$ \Msun, while the most massive has $m_{\rm SP} = 61$ \Msun\ by
the end of the simulation. 

We also use the non-feedback counterpart of LAF1 simulation, labeled
LAF0 in Paper I, as a `control simulation' to test feedback
effects. In this simulation, the most massive SP has $m_{\rm SP}
\approx 100$ \Msun\ at its final time. As the SPs in our simulations
have stellar masses, hereafter we will refer to them simply as
`stars'.

\subsubsection{Feedback scheme}
\label{feedback}
For simplicity and cleanliness, in our simulation with feedback, we
only consider a sub-grid model of UV photoionising radiation from
massive stars. This choice is partially justified by the fact that a
number of studies have concluded that this form of feedback is likely
to be the dominant form of energy injection into giant molecular
clouds (GMCs) leading to their dispersal \citep[e.g.,] [] {Matzner02,
  Haid+19}.  Nevertheless, in a future work we plan to test other
feedback sources.

For computing the ionising feedback from massive
stars, in this work we first determine the size of an \hii\ region around
the massive stars. To do that, we compute the \citet{Stromgren39}
radius corresponding to the star's mass as

\begin{equation}
R_{S} \equiv \left( \frac{3}{4\pi}\frac{S(M_{\star})}{\alpha
n^{2}_{\rm LOS}}\right)^{1/3}, 
\end{equation}
with $S(M_{\star})$ the ionising flux produced by a star of mass
$M_\star$, $\alpha = 3.0 \times 10^{-13} {\rm cm}^{3} {\rm s}^{-1}$
the hydrogen recombination coefficient and $n_{\rm LOS}$ a uniform
line-of-sight (LOS) characteristic density in the \hii\ region computed
as the geometric mean of the density at the site of the star, $n_{\rm
SP}$, and the density at the test grid cell, $n_{\rm tc}$, so that
$n_{\rm LOS} = \sqrt{n_{\rm SP}n_{\rm tc}}$. In Paper I it was shown
that this simplified prescription reproduces correctly the expansion
of an \hii\ region, although it neglects the effect of ``shadows''
behind clumps. Nevertheless, our results are consistent, for example
in measured global star formation efficiencies, with those obtained
from simulations with full radiative transfer implemented
\citep[e.g.,] [] {ZA+19}. For $S(M_{\star})$ we use tabulated data
provided by \citet*{DM+98}. We assume that only stars with masses
greater than 1.9 \Msun\ inject any significant feedback into the ISM.

Once $R_{S}$ is computed for each line of sight, we
assign a temperature of $10^{4}$ K to all cells whose distances $d_{\rm LOS}$ to
the star satisfy $d_{\rm LOS} < R_{\rm S, LOS}$, and we turn off
the cooling in these cells to prevent very
dense cells from radiating their thermal energy too quickly. 
This thermal state will last for a
time $t_{s}$ which depends on the mass of the star according to:
\begin{equation}
 t_{s} = \left\{
  \begin{array}{lr}
   2\; {\rm Myr} & \text{if }  m_{*} \leq 8 {\rm M}_{\odot} \\
   222\; {\rm Myr} \left(\frac{m_{*}}{{\rm M}_{\odot}} \right)^{-0.95} & \text{if }  m_{*} > 8 {\rm M}_{\odot} 
   \end{array}
   \right.
   \label{eqtime}
\end{equation}

For stars with $m_{*}$ lower than 8 \Msun, this time characterizes
their stellar wind phase, while for stars with $m_{*}$ greater than 8
\Msun, $t_{s}$ is a fit to their lifetimes as given by
\citet{Bressan+93}. Note that with this feedback prescription we are
accounting for non-local radiation effects: ionising photons from
massive stars affect not only the parent cell where a star was born,
but all the nearby cells within $R_{S}$. However, our single-fluid
prescription does not follow the neutral and ionised components. 
As we mentioned above, we simply set the temperature of cells 
that are within the $R_{\rm S, LOS}$ from an ionising star to $10^{4}$ K,
and then let them return to their thermal-equilibrium ($\Gamma = n \Lambda$) 
temperature once the star ``turns off'' according to the stellar lifetime 
given by eq.\ (\ref{eqtime}).

\subsection{Procedure}
\label{proc}

A number of SCs were formed in the simulated box (see Paper II for its
general description). In this work we select the G1--2 group to show
the effect of the feedback on the gas distribution and on the stars
around the SC. For simplicity, henceforth we will refer to G1--2 group
as G2. This was born in a region clearly detached from other star
forming regions (several tens of parsecs away) and allows
us to start the analysis with an unambiguous membership identification
of the stars. Note that other regions in the simulation start forming stars 
at almost the same time ($\pm$ 1 Myr) than G2; thus, ionising radiation
from their massive stars does not affect the evolution of G2.  
Within this region there are approximately 390 \Msun\ in stars in the
 LAF1 simulation by the end of the last analysed snapshot, while in the 
 LAF0 run, around 883 \Msun\ were converted into stars.
Another filamentary-like star forming region which appeared in the LAF1
simulation will be the subject of a future analysis.

We will use the center of mass of the stellar cluster, denoted ${\rm
CM}_{\rm s}$, as the origin of the reference frame to compute its properties. As the stars are the source
of photoionisation feedback, the center of mass of the stellar
distribution represents the natural framework to compute its effect on
the surrounding gas distribution. In Paper I we made an effort to
determine, at each snapshot in the simulation, the membership of stars
to the G2 cluster disregarding runaway stars to update ${\rm CM}_{\rm
s}$ in subsequent time steps. Instead, for simplicity, here we include
in our analysis all the stars that were born clearly within the star
forming region first identified at $t \approx 19$ Myr in the LAF1
simulation, and follow their evolution over $\approx$ 10 Myr.
During this period, a number of small sub-units
appear in the surroundings of G2. We easily identify these
subgroups by eye, and we follow their evolution by tracking the positions of their
stars. 

Similarly to what is done for G2, 
the center of mass is also computed for each subgroup, and all the 
stars born within the radius determined by the outermost star are
considered as a new stars belonging to the corresponding subgroup. 
Obviously, some complications arise when groups start to merge. 
When that is the case, the center of mass is now computed 
considering all the stars from the merged groups. We will discuss this
with more detail in Sec. \ref{stars}.

Given the center of mass of
the stellar distribution, we compute several properties of the gas and
stars within different radii to analyze the imprint of stellar
feedback on them. Thus, we are not constraining our analysis to a
particular region, such as the region within the radius of the group,
since we do not attempt to define such radius. Obviously, the center of
mass depends on the stars considered but, given our
approach to measure the cluster properties at different radius, and
the fact that the differences on the location of the center of mass
when we exclude runaway stars are minimal,\footnote{The distance
between the center of mass that include the runaway stars (our
fiducial approach) and the center of mass computed excluding such
stars is within the innermost radius used in this work to compute
stellar and gas properties at all times during the 10 Myr of evolution
of the G2 cluster.} we do not expect that avoiding the definition of
a radius group shall affect our results.

\section{Results}
\label{results}

\subsection{General evolution of the simulation} \label{Sevo}

In this section we provide a brief description of the evolution of the
simulation that leads to the formation of the clusters studied. For
further details, the reader is referred to \citet{Colin+13} and
\citet{VS17}.

The colliding streams form a cold dense layer through a shock and a
condensation front \citep{KI00, VS+06}. The layer becomes turbulent by a
combination of various instabilities \citep{Vishniac94, KI02, KI04,
  Audit+05, Heitsch+06, Inoue+06}, with the turbulence being
moderately supersonic with respect to the cold gas. However, due to
the added turbuelent velocity field in the initial conditions, the
layer is strongly bent. Furthermore, as noted in Sec.\ \ref{method},
the streams have a length of 112 pc, and collide in the central $(y,z)$
plane of the simulation. That is, the streams have a finite length,
and are completely contained within the 256-pc numerical box. At their
inwards speed of 5.9 \kms, they complete their collision after 18.7
Myr. 

During this time, the layer has been increasing its thickness and
column density at roughly constant mean volume density, until it
exceeds its Jeans mass and begins to contract gravitationally.  Due to
the turbulent motions, however, the contraction is highly
non-homologous and anisotropic, forming filaments that accrete from
the cloud and funnel the gas to the collapse centers, in a river-like
flow \citep{GV14, VS+19}. Moreover, when the filaments have increased
their linear mass density sufficiently, they undergo local low-mass
collapse that constitute secondary star-formation sites \citep{GV14}
in the periphery of the main collapse centers, usually referred to as
``hubs'', in a ``conveyor-belt''-like fashion \citep{Longmore+14}.

Within this context, the first stars appear in the simulation at an
absolute time (i.e., from the start of the simulation) $\sim 19$ Myr.
In what follows, we analyse the properties of two stellar clusters
that begin forming at this time. For simplicity, we describe
  the evolution in terms of the time measured after the onset of star
  formation ($t_{\rm SF}$) in this region, which we refer to as $\tage =
  t-t_{\rm SF}$.

\subsection{Indirect effect of the feedback on the stars via the gas}
\label{Sdistro} 

\begin{figure*} 
\hspace{0.3cm}\includegraphics[width=0.9\textwidth, height=20cm]{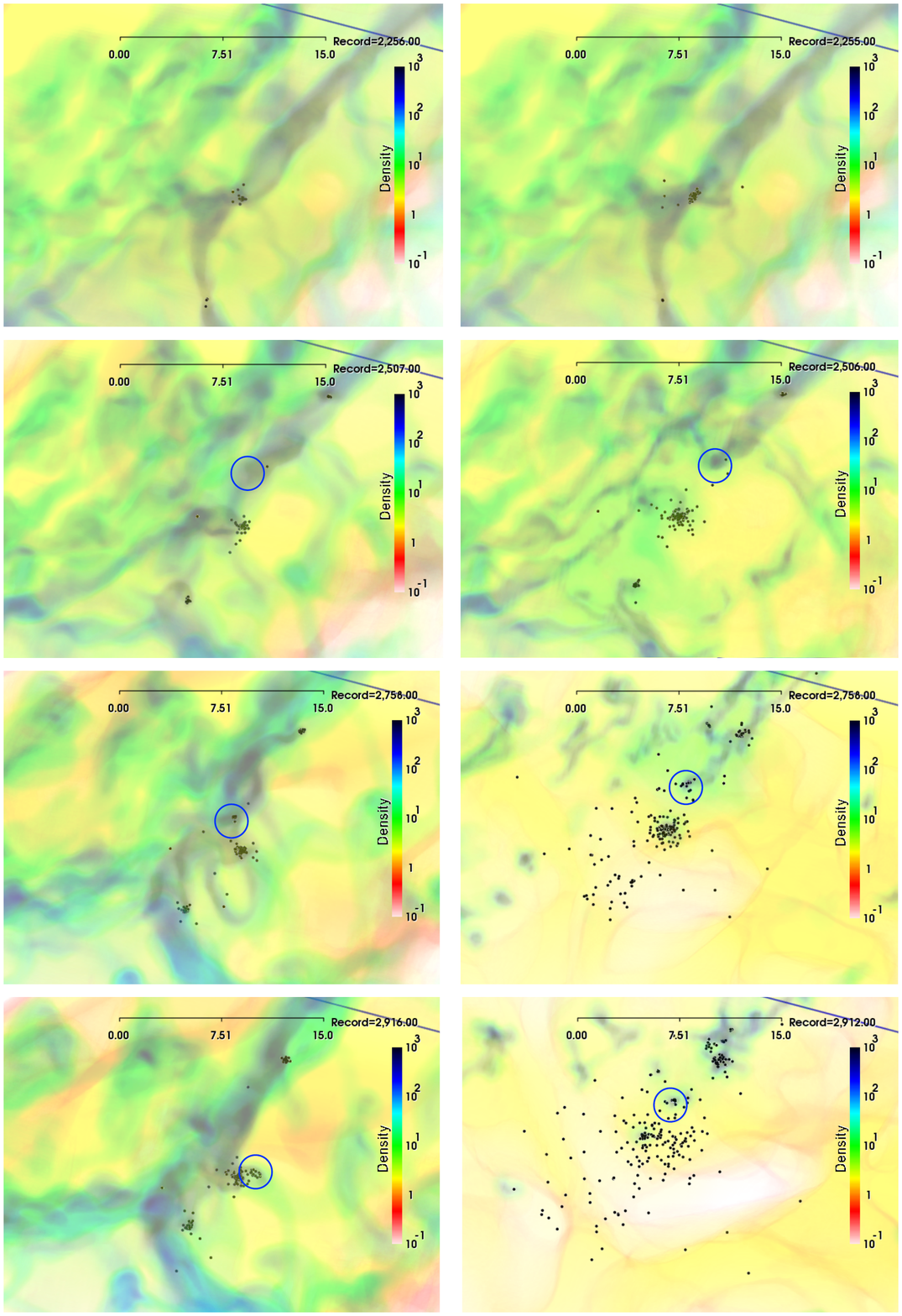} 
\caption{Non-Feedback (left panels) vs Feedback run (right panels)
  comparison of the G2 cluster in the simulated box at times $t =
  22.5, 25.0, 27.5$ and $29.1$ Myr since the start of the simulations.
  These times correspond to $t_{\rm age} \approx 3.7, 6.2, 8.7$ and
  $10.2$ Myr since the formation of the first star in the group. The
  color bar indicates the gas number density in $\pcc$. 
      Note that the lower-density gas has a higher transparency in the
      rendering, and thus it is not very prominent.
  The ruler indicates a scale of 15 pc. Black dots represent
  individual stars born within the densest gas clumps. Ionising
  feedback from massive stars sweeps the gas out of the star forming
  region showing ring-shaped features in the gas distribution $\approx
  5$ Myr after the formation of the stellar group (second panel from
  top at the right) and clear the region from gas after $\approx 10$
  Myr (bottom panel at the right). Ionising feedback has a strong
  effect too in the spatial stellar distribution: the simulation
  including feedback forms more extended stellar clusters than the one
  without feedback.  The circles indicate the position of a
  clump/stellar group that in the run without feedback (left panels)
  undergoes a close encounter with the stellar group in the main hub,
  but is pushed sideways by the expanding shell around the main hub in
  the run with feedback and thus avoids the close encounter.}
\label{fig1}
\end{figure*}

In Fig. \ref{fig1} we show the spatial stellar distribution of the
simulated cluster G2 in the runs without (LAF0, left panels) and with
feedback (LAF1, right panels) at four different times (from top to bottom,
$t = 22.5, 25.0, 27.5$ and $29.1$ Myr since the start of the
simulations). These correspond to times $t_{\rm age} \approx 3.7, 6.2,
8.7$ and $10.2$ Myr since the formation of the first star in the
group.  The gas density is color coded in the figure according to the
color bar in each panel.  

Fig.\ \ref{fig1} directly illustrates one of our main results: once
massive stars are born, feedback becomes dominant and not only sweeps
the gas out of the stellar cluster $\sim 5$ Myr after the formation of
its first star (see the right panel in the second line in Fig.
\ref{fig1} where an \hii\-like region can be identified), but it also
strongly affects the spatial distribution of the stars, which is much
more extended by the final time in the run with feedback (see the
right panel at the bottom in Fig. \ref{fig1}) than in the run without
feedback.  In the latter, the cluster is tightly concentrated within a
region of $\approx 2$ pc of radius, with only a few smaller subunits
appearing in the surroundings, which are clearly
detached from the main cluster. Thus, feedback not only
disperses the gas, but induces a more scattered initial configuration
of the cluster which is not only due to the reduction in the
gravitational potential of the region once the gas is removed. In
addition, the gas being pushed away from the main hub continues to
form stars for some time, and therefore those stars form at more
distant locations from the CM than in the case without feedback. For
example, in Fig.\ \ref{fig1}, the circles indicate the position of a
clump/stellar group that in the run without feedback (left panels)
undergoes a close encounter with the stellar group in the main hub,
orbiting around it and producing a tidal tail of stars. The same clump
in the run with feedback is pushed sideways by the expanding shell
around the main hub, and thus never undergoes the close encounter with
the hub and its stars.

The effect of the feedback is also manifest in that, in the non-feedback
run LAF0, after 10 Myr of evolution since the formation of the first star
(bottom left panel), the accretion along the filament onto the
main hub continues, with the filament being as dense and well defined
as at the start. Instead, in run LAF1, the dense gas has
been almost completely evacuated in the neighbourhood of the cluster,
with only traces of gas being left at the sites of secondary
cores/groups.

Finally, the well-known effect of the stellar cluster expanding as the
gas is removed due to the loss of the latter's gravitational potential
\citep[e.g.,] [] {Tutukov78, Lada+84, Baumgardt+07, Parmentier+08}
is also clearly seen in Fig.\ \ref{fig1}, which shows that the stellar
distribution in the run with feedback is much more extended than that
in the run with no feedback. In this case, the stellar expansion
begins to be clearly noticeable $\sim 8$ Myr after the formation of
the first star.

\begin{figure*} 
\includegraphics[scale=0.34]{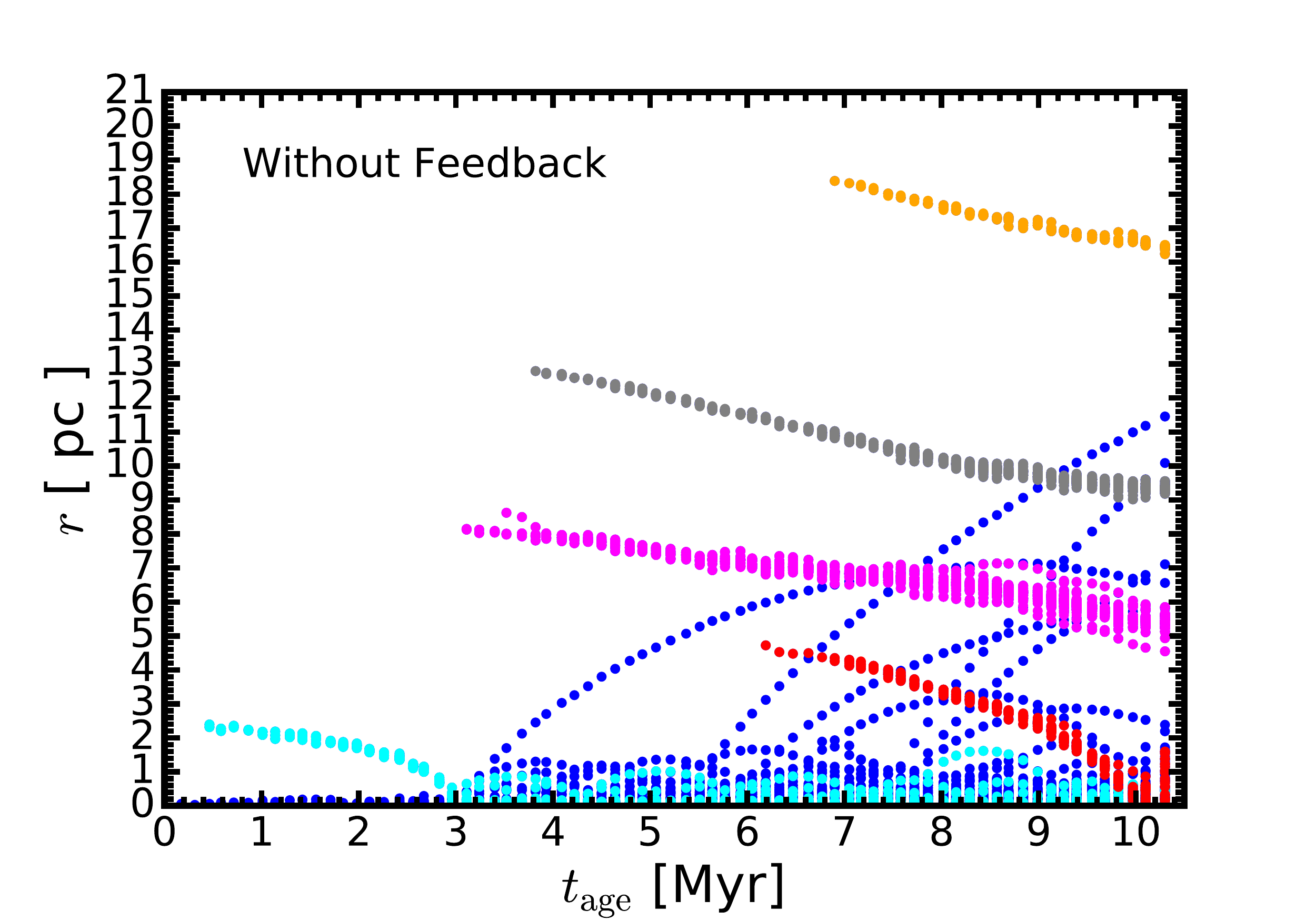} 
\includegraphics[scale=0.34]{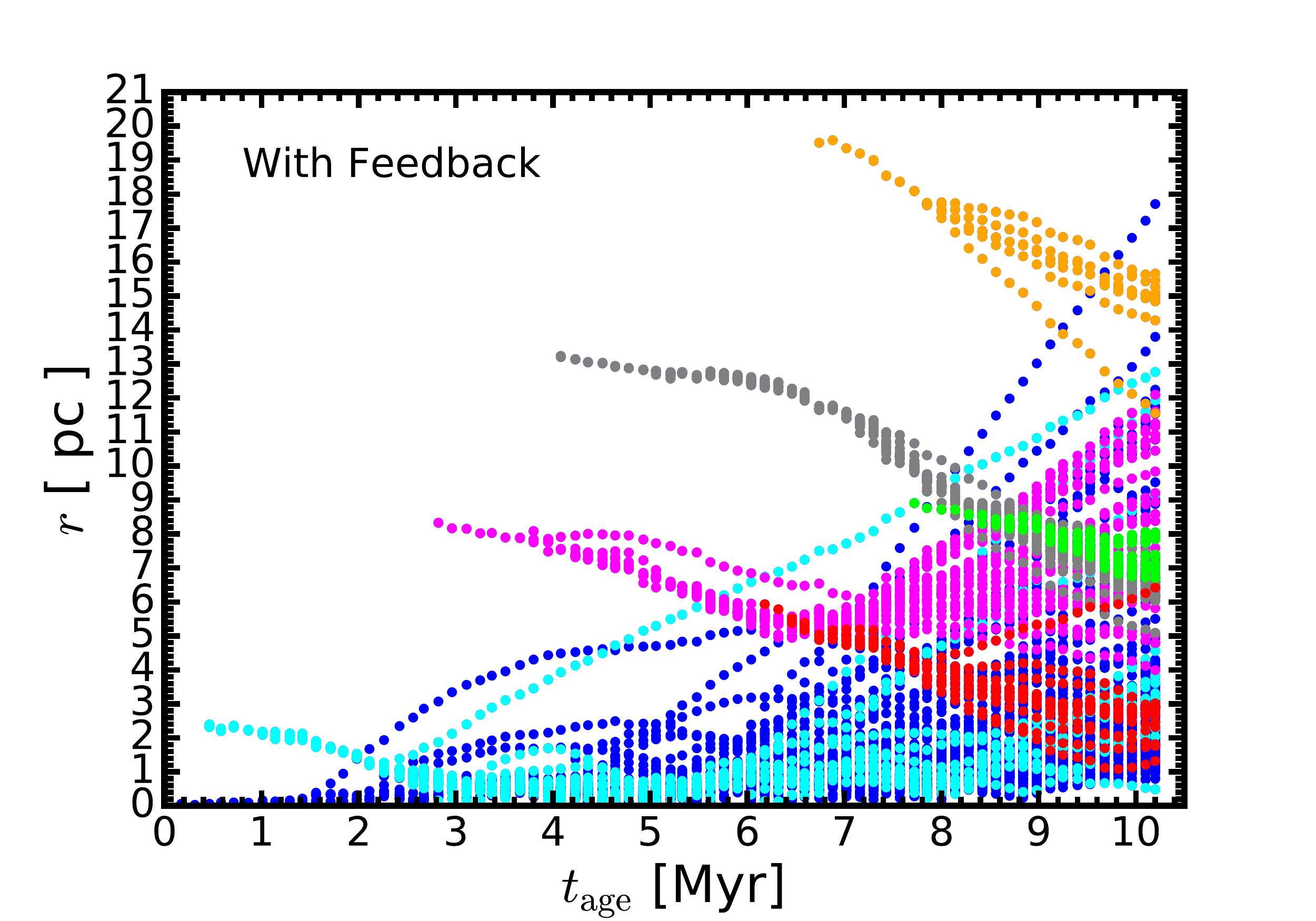}
\caption{Stellar particle distribution as a function of $t_{\rm age}$,
  the elapsed time since the start of the formation of the cluster,
  for the run without feedback (LAF0, left panel) and the run with
  feedback (LAF1, right panel). The dots represent individual stars,
  and are colored according to their membership to the various
  subgroups, subG1 (cyan dots), subG2 (magenta dots), subG3 (grey
  dots), subG4 (orang dots), subG5 (red dots) and subG6 (green dots,
  only in the right panel). The stellar groups formed in run LAF0 are
  extremely concentrated (left panel), so some of them never merge
  with the others, while all groups merge in run LAF1, indicating that
  feedback promotes the mixing of the stellar populations by affecting
  their kinematic properties (right panel). Note also that one
  subgroup (green dots) forms at late times in run LAF1 that does not
  have a counterpart in run LAF0, constituting one (somewhat
  exceptional) instance of feedback-triggered star formation. }
\label{groups}
\end{figure*}

\subsection{Masses of the most massive stars}
\label{sec:most_massive} 

The truncation of the gas supply along the filaments in the feedback
run, LAF1, has the additional consequence of limiting the masses of
the most massive stars that form. Specifically, the most massive star
formed in the whole numerical box of run LAF1 has 61 \Msun, while in
run LAF0 it reaches 100 \Msun. If we consider only the star forming
region around G2, the most massive star in LAF1, formed $\approx 5$
Myr after SF began, has $\sim 30$ \Msun, while the most massive star
in run LAF0 has $\approx 60$ \Msun, and forms around 8 Myr after the
onset of SF. A stellar particle with a mass similar to that of the
most massive star in LAF1 is formed more than a Myr before in run
LAF0. Thus, feedback delays even more the formation of massive stars
by controlling the accretion onto the star-forming hubs and clumps.

\subsection{Hierarchical cluster assembly}
\label{stars}

The formation of smaller subunits or subgroups occurs in both runs,
but as feedback ``inflates'' the stellar distribution, the subunits
merge more rapidly in run LAF1, while they remain more
clearly separate in run LAF0. Nevertheless, traces of their
separate origins are still seen as local density enhancements in the
resulting cluster. Also, it is important to note that the different
secondary sites begin forming stars at different times. This is shown in
Fig. \ref{groups}, where we plot the 3D radial distance $r$ of each
star (colored points) with respect to the centre of mass of the G2
cluster (points in blue) as a function of the time elapsed since the
formation of the first star, $t_{\rm age}$. Only stars that were born
within the G2 cluster at each time were used to compute the centre of
mass. Thus, we explicitly exclude stars that form in a clearly
different gas clump, even if they eventually merge with the G2
population.

As is shown in the right panel of Fig. \ref{groups}, six subgroups
form in the simulations that end up comprising group G2. We will refer
to them as subG1 (cyan dots), subG2 (magenta dots), subG3 (grey dots),
subG4 (orang dots), subG5 (red dots) and subG6 (green dots). In run
LAF1, subgroups subg1 and subG5 fall onto G2 some time after their
formation, while subgroups subG2, subG3 and subG6 are close enough to
G2, that at late times, stars formed within these subgroups mix with
the G2 population, without properly ``falling'' onto it. And
viceversa, some stars from the central hub have been ejected, so at late times they
have reached the locations of these subgroups. Subg4 is the outermost
subgroup formed in this region of the numerical box, appearing $\sim
20$ pc away from the centre of mass of G2. Thus, we define the star
forming region around G2 (SFRegG2) as the one that encloses all these
subgroups; basically a 20-pc sphere around G2. This is somewhat
arbitrary, but we will use it just to characterize the outskirts of
G2, and our results do not depend sensitively on this choice. 
SFRegG2 manages to gather a mass $\la 10^{4}$\Msun\ 
in dense gas during its evolution; thus, 
our results should be compared to observational or numerical inferences
obtained for MCs in this mass range.

Most of the subgroups form in small (``secondary'') collapsing sites
within the filaments of dense gas that feed the main cluster G2. They
started as small groups of stars confined to a small area, but a few
megayears later they are dispersed due to the effect of feedback,
either local, or from the main hub. SubG2 is the extreme case of this
situation: as soon as feedback disperses the gas around G2, it also
affects the stars constituting subG2, practically destroying it within
a few megayears.  Thus, in general, {\it feedback from massive stars
  causes an expansion of the stars and, as a result, it facilitates
  the mixing of populations from different subgroups}. Thus, possible
``primordial'' age gradients \citep{VS17} might be blurred by this
process.  It is easy to verify that feedback from massive stars is the
cause of this effect by comparing the left panel (run without
feedback) with the right one (fiducial run including feedback) of Fig.
\ref{groups}: the subgroups form in both simulations, but are more
tightly concentrated in run LAF0 in comparison to run LAF1. Thus, their
populations never mix in LAF0, contrary to what occurs in LAF1
(compare the radial extents of each group in the two panels of Fig.\
\ref{groups}).

\begin{figure*} 
\hspace{-0.2cm}
\includegraphics[scale=0.24]{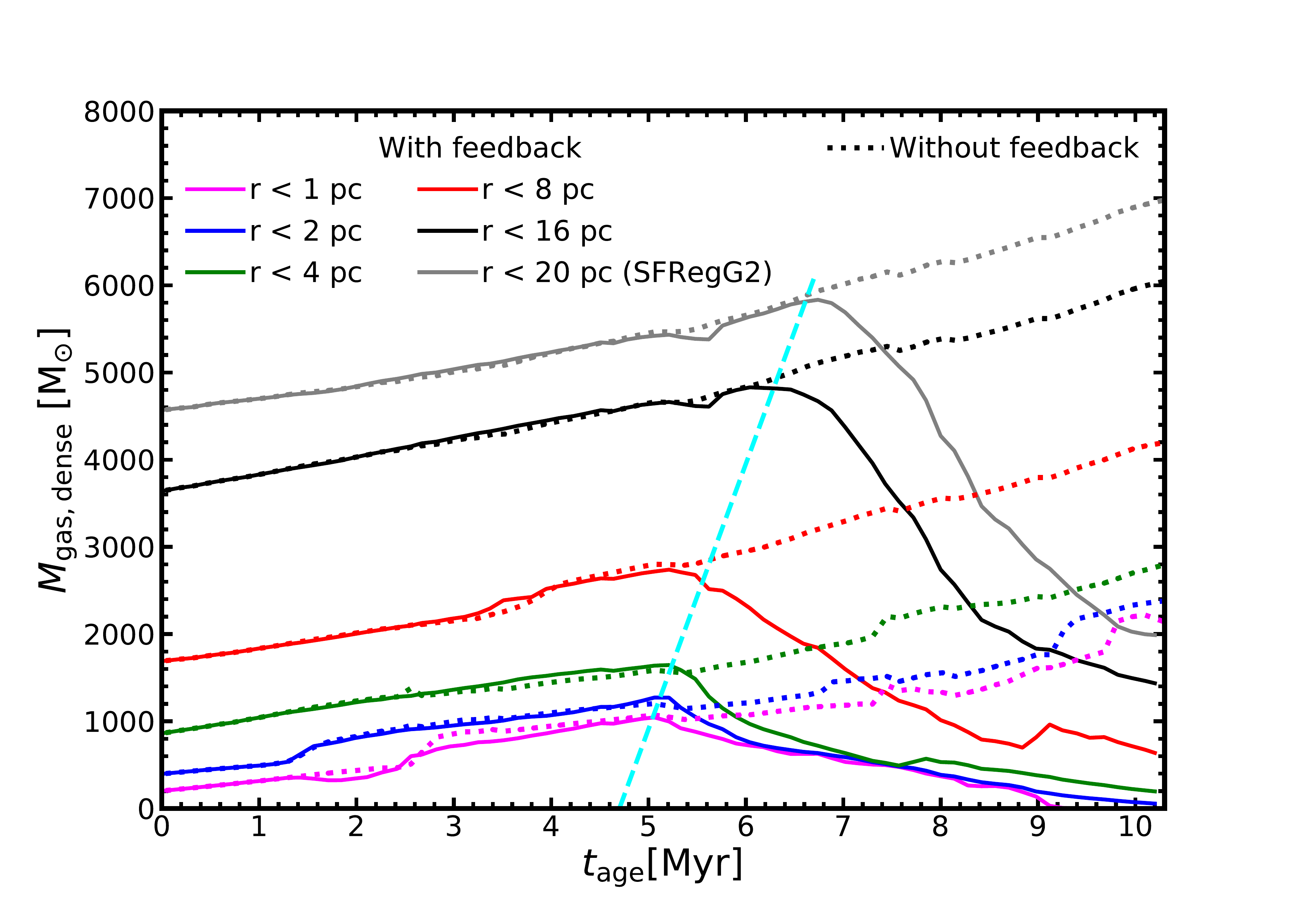} 
\includegraphics[scale=0.24]{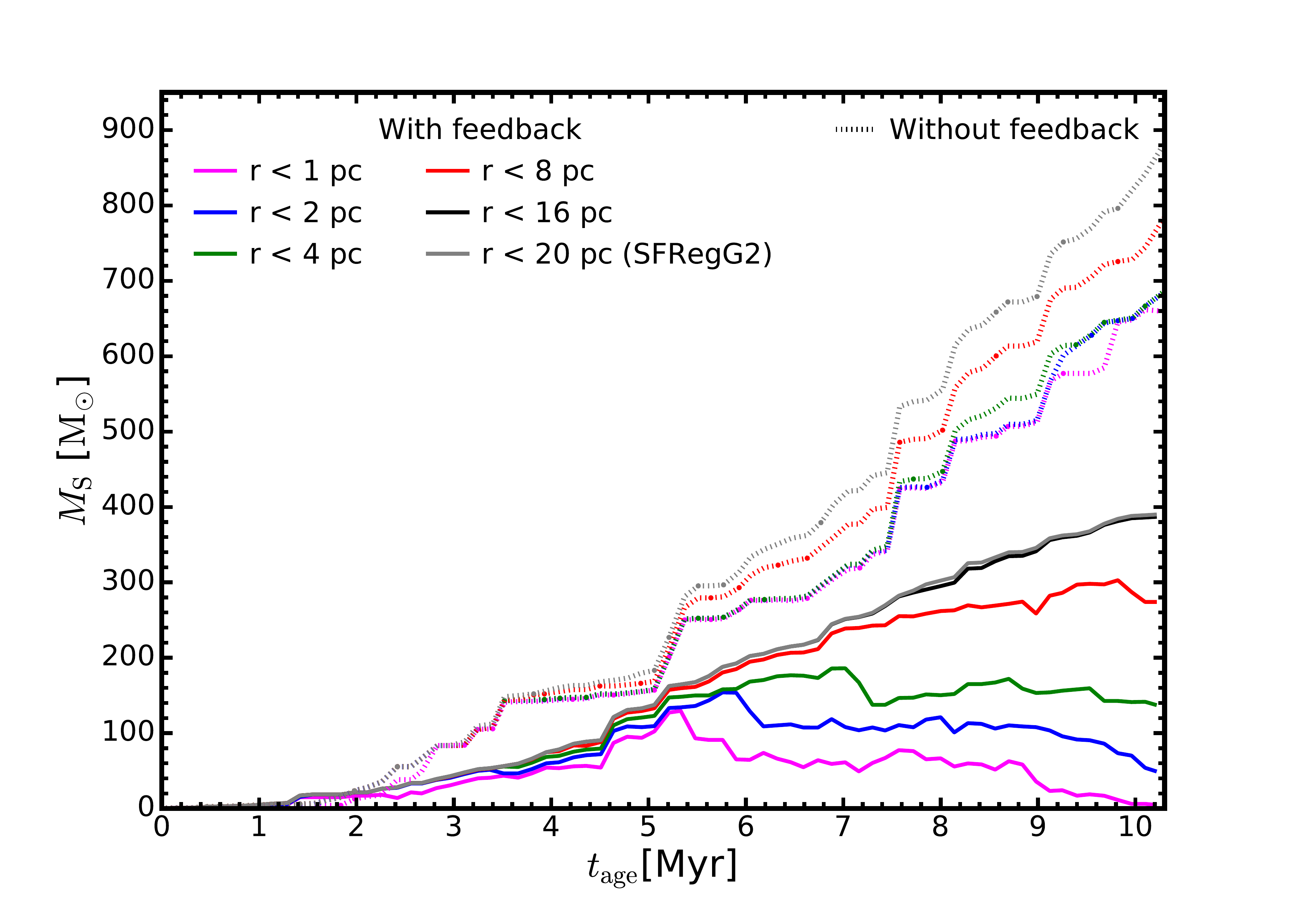} 
\caption{{\it Left:} Evolution of the dense gas mass enclosed within
  radii $r = 1, 2, 4, 8, 16, 20$ pc from the center of mass of G2. The
  solid and dotted lines respectively show the results for the
  simulations with (LAF1) and without (LAF0) feedback. The effect of
  feedback from massive stars in the dense gas mass distribution is
  clear at $\approx 5$ Myr after the formation of the first star
  ($t=0$ in this figure). The cyan dashed
    line in the left panel represents
    the times at which an expanding front moving at a speed of 11 \kms\
    intersects the various lines. {\it Right:} Evolution of the
  stellar mass in the same spherical regions as in the left panel. The
  stellar mass is seen to stop increasing or even decrease in the run
  with feedback, while it continues to increase at an accelerated pace
  in the run without feedback.}
\label{fgasdense}
\end{figure*}

\subsection{Dense gas consumption or removal?}
\label{gas}

The evolution of the SCs discussed above depends fundamentally on the
amount of dense gas available to form stars at a given time. This
reservoir depends on the accretion rate into the region, consumption
to form stars (quantified by the SFR) and dispersal or evaporation due
to feedback. In this work, we do not distinguish between dispersal and
evaporation, and instead consider them together,
referring to them generically as gas {\it removal}.
 
In order to determine how much gas is available to form stars as a
function of time, the left panel of Fig.\ \ref{fgasdense} shows the
amount of dense gas ($\ndens > 100\ \pcc$) enclosed within spheres of
radius $r = 1, 2, 4, 8, 16,$ and 20 pc from the center of mass of G2
for the two simulations. For comparison, the right panel of this
figure shows the mass in stars within the same spherical regions.

As shown by the solid curves in the left panel of Fig.\
\ref{fgasdense} (which denote the feedback run, LAF1), at time $t_{\rm
  age} \sim 5$ Myr, when the first very massive ($M = 30$ \Msun) star
forms,\footnote{Before that time, the most massive star in the region
  had a mass $M=4.1$ \Msun.} an important qualitative change in the
behaviour of the feedback simulation occurs: before that time, the
mass in dense gas had been consistently increasing throughout region
SFRegG2.  However, after that time, the dense gas mass begins to
decrease until there is no more dense gas in the innermost parts of
SFregG2 at $t_{\rm age} \approx 10$ Myr. Thus, the gas is either fully
consumed by star formation or fully removed by feedback within a
radius of $\sim 4$ pc over a span of another $\sim 5$ Myr. On the
other hand, in the non-feedback run LAF0, the dense gas mass simply
continues to increase within all of the radii considered.

Note that the onset of the decrease of the dense gas mass occurs at
later times for spheres of larger radii (see, for example, the curves
for $r = 16$ and 20 pc---black and grey curves, respectively---in
Fig.\ \ref{fgasdense}). This is indicative of the expansion of the \hii\
region around the main hub, and implies a velocity $\approx 11$
\kms. It is important to note that the massive stars responsible for
the ionising flux in the region were born just before the dense gas mass
begins to decrease in the central parsec. That is, there is a time
during which the star-forming region can accrete and grow essentially
unimpeded, before the massive stars form. After this, the region is
dispersed essentially in the crossing time of the sound speed in the
ionised gas.

We define the dense gas mass removal rate from sphere of radius $r$ as
\begin{equation}
\begin{aligned}
\dot M_{\rm g, dense} & = - \frac{\Delta M_{\rm g, dense}(r)}{\Delta t} +
\frac{\Delta M_{*}(r)}{\Delta t} \\ 
 &  \equiv - \frac{\Delta M_{\rm g, dense}(r)}{\Delta t} + {\rm SFR(r), }
\end{aligned}
\label{eqflux}
\end{equation}
where the masses are computed inside radius $r$, 
$\Delta t$ is the elapsed time between two
consecutive snapshots in the simulation ($\sim 0.14$ Myr), $M_*(r)$ is
the mass in stars within radius $r$, the changes in the masses are
computed over the time interval $\Delta t$, and the second equality
defines the star formation rate within radius $r$, SFR$(r)$.  By
including the instantaneous SFR in the balance to compute $\dot M_{\rm
g,dense}$ we account for gas consumption by conversion to stars.  This
also avoids spurious contributions by previously existing stars that
enter or leave the spherical region.

\begin{figure*} 
\includegraphics[scale=0.31]{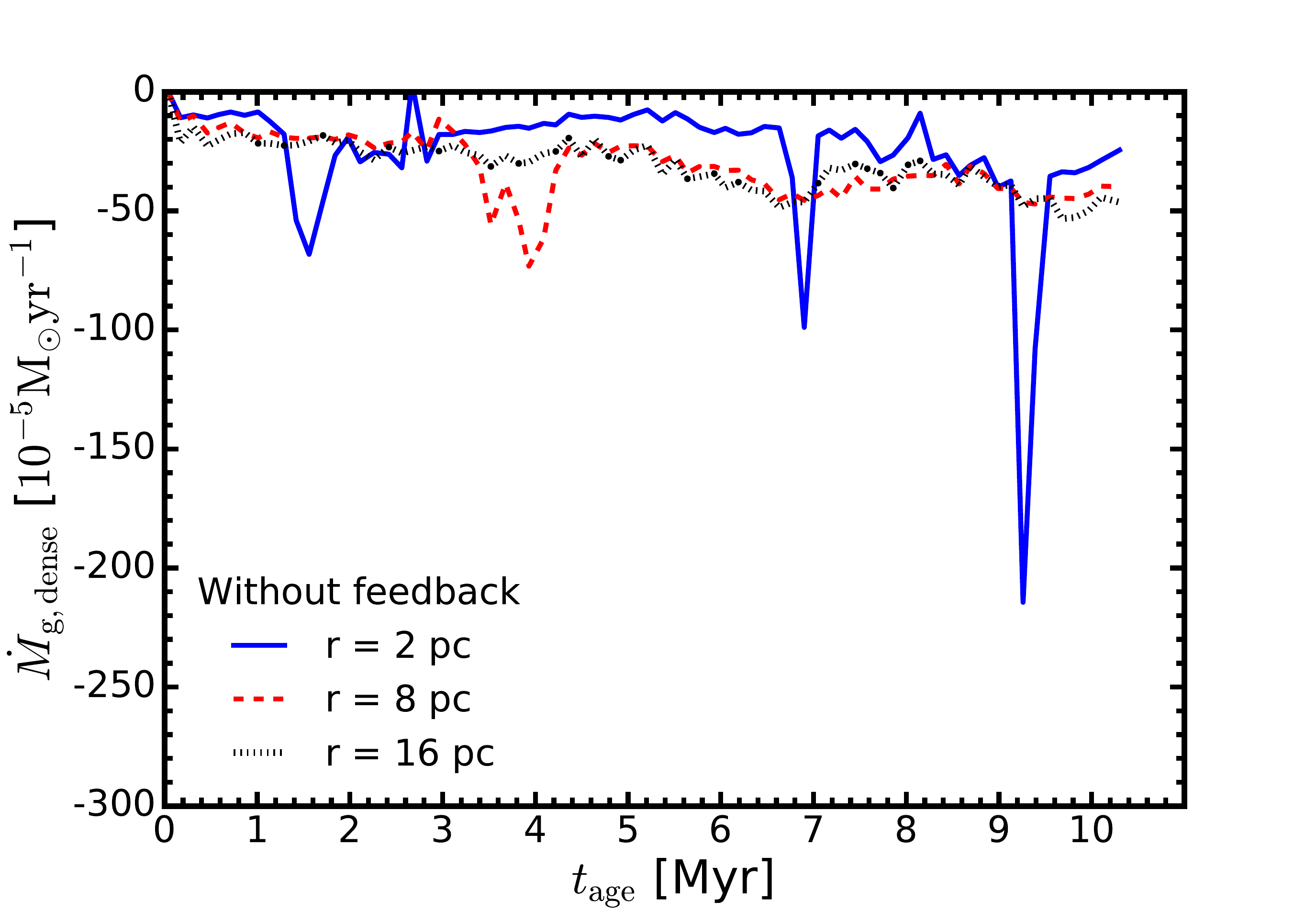} 
\includegraphics[scale=0.31]{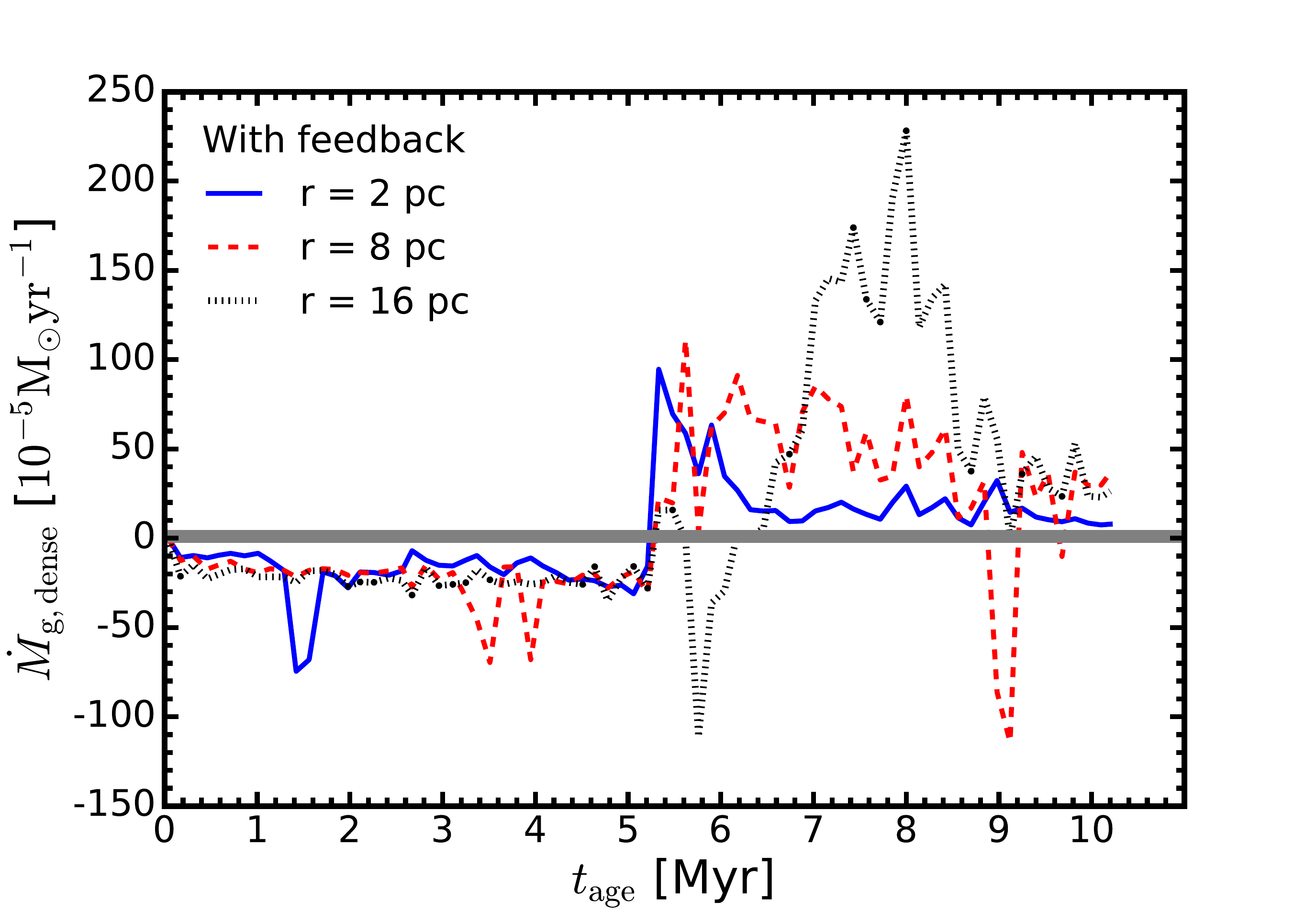} 
  \caption{Evolution of the dense gas mass rate of change $\dot
  M_{\rm{\rm g,dense}}$ within spheres of radius $r = $ 2 pc, 8 pc and
  16 pc. Run LAF1, with  feedback (right panel), shows two
  clear stages during the first 10 Myr of stellar cluster evolution:
  a strong collapse stage at the beginning ($\dot M_{\rm{\rm
  g,dense}} < 0 $, $\it{gas\ accretion}$), and a gas removal stage
  once stellar feedback becomes important ($\dot M_{\rm{\rm
  g,dense}} > 0 $).  }
\label{flux}
\end{figure*}

As defined by Eq.\ (\ref{eqflux}), negative values of $\dot M_{\rm g,
dense}$ imply that accretion is more important than dispersion and
star formation. The left panel of Fig.\ \ref{flux} shows that, in the
non-feedback run LAF0, $\dot M_{\rm g,dense}$ is always negative,
meaning that {\it the region always accretes faster than it can
convert gas to stars}, explaining the continued dense mass increase at
all radii seen in Fig.\ \ref{fgasdense}. Instead, in run LAF1 (right
panel of Fig. \ref{flux}), the two stages are again clearly
identified. In the first, accretion of dense gas predominates, until
the time when sufficiently massive stars appear and feedback begins to
dominate, defining the beginning of the second stage, represented by
positive values of $\dot M_{\rm g,dense}$, in which dense gas removal
dominates over accretion. Note that this does not mean that accretion
is halted instantaneously, but only that the balance between
accretion and removal begins to be reversed.

\subsection{Evolution of the star formation rate}
\label{Sfr}
\begin{figure} 
\hspace{-0.4cm}
\vspace{-2.0cm}
\includegraphics[scale=0.26]{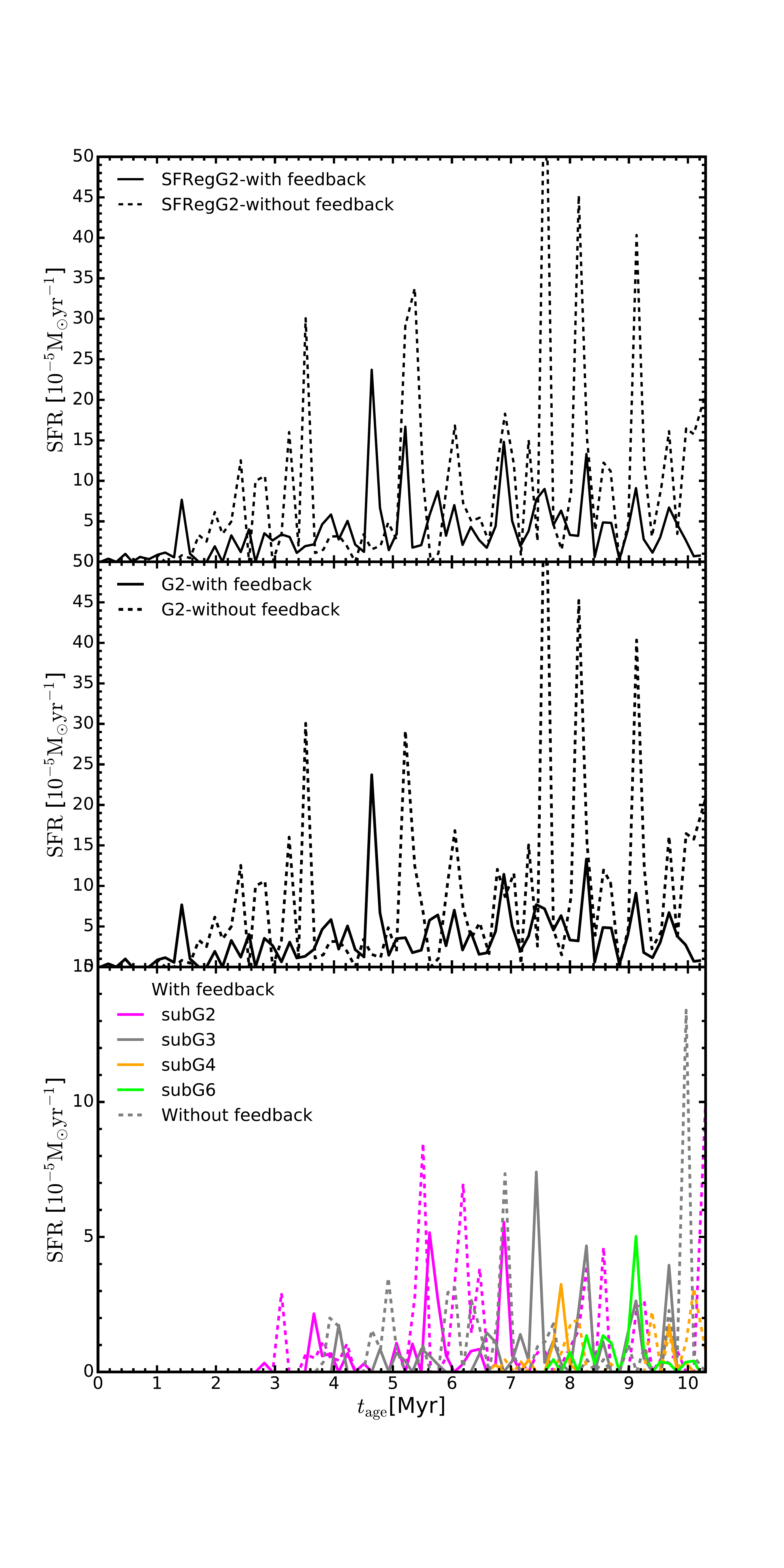} 
  \caption{Instantaneous star formation rate \sfrt\ as a function of
  time for (from top to bottom): the star forming region around G2
  (SFRegG2, $\rm{r} < $ 20pc), the G2 group, and for the four
  subgroups formed farthest from G2 (see Figure \ref{groups}). It does
  not matter if we consider the whole region around G2 (SFRegG2) or
  just the inner four parsecs around G2, \sfrt\ increases with time
  peaking $\sim$5 Myr after the first star was formed, then it decreases;
  and clearly the non-feedback run (dashed lines) are forming stars at
  a higher rate than the respective regions in the run that include
  ionising feedback from massive stars (solid lines). \sfrt\ for the
  subgroups (bottom panel) shows that sometimes feedback just stops
  the formation of stars (see subG2 and the respective curve for the
  simulation without feedback), and sometimes it triggers SF, as is
  the case for subG6, which does not even have a counterpart in the
  simulation without feedback.}
\label{sfr}
\end{figure}

The effect of the feedback on the SFR is one of the fundamental
questions in the problem of how star formation self-regulates.  In the
top panel of Fig. \ref{sfr} we show the evolution of the star
formation rate as defined by the second term of the right-hand side of
eq.\ (\ref{eqflux}), using a sufficiently large radius $r$ to
encompass all of the star formation in region SFRegG2 in the two
simulations. In both runs, the SFR is extremely intermittent (bursty)
over the entire evolution. However, in the feedback run LAF1, the SFR
first increases with time, reaching a maximum at $\tage \approx 4.5$
Myr, and then it begins to decrease on average, until dense gas is
depleted and star formation is fully quenched in the region. Instead,
in the non-feedback run LAF0, as long as gas accretion goes on, the
SFR continues to grow in time, and reaches higher values than in run
LAF1. By the final time shown, $\tage \sim 10$ Myr, the mass in stars
in region SFRegG2 in run LAF0 is $\sim 2.3$ times larger than the
corresponding mass in run LAF1 and is still increasing (see the right
panel of Fig.\ \ref{fgasdense}), while it has stopped incrreasing in
run LAF1, and the SFR has shut off (Fig.\ \ref{sfr}). The situation is essentially the same for the central group
G2 only (middle panel in Fig.\ \ref{sfr}). This is because most of the
star formation activity in the whole SFRegG2 occurs in the clump that
gives rise to G2. Specifically, $\sim 87\%$ and 91\% of the total mass
in stars within SFRegG2 were formed in G2 after 10 Myr of evolution in
runs LAF1 and LAF0, respectively.

The difference in stellar mass between the two runs (a factor $\sim
2.3$) by the final times ($\tage \sim 10$ Myr) shown in Fig.\
\ref{fgasdense} may seem too low to provide any effective
self-regulation of star formation.  However, what happens is that, in
the non-feedback run, star formation continues for a long time after
the last times shown in this figure, as can be inferred by the fact
that the stellar mass is not only still increasing in run LAF0 at that
time, but it is doing so at an accelerated pace. Instead, SF has been
almost completely quenched by the feedback at this time in run LAF1,
and no new stars are forming in the region. But, comparing the gas and
stellar masses at $\tage = 7$ Myr, at which these masses are still
similar in the two simulations, we see that the instantaneous star
formation efficiency,
\begin{equation}
{\rm SFE}(t) = \frac{M_{\rm S}(t)} {M_{\rm dense}(t) + M_{\rm S}(t)},
\label{eq:sfe}
\end{equation}
is $\approx 4\%$ in LAF1 and $\approx 6.5\%$ in LAF0, since $M_{\rm
dense} \approx 5800 M_\odot$ in both runs, while $M_{\rm S} \approx
250 M_\odot$ in LAF1 and $M_{\rm S} \approx 400 M_\odot$ in LAF0. So,
at this time, the SFE is still quite low in the two simulations, and
consistent with observations. This is due to the multi-timescale
nature of the hierarchical collapse and the corresponding acceleration
of SF \citep{VS+19}. In order for the non-feedback run LAF0 to more
thoroughly convert its gas to stars, significantly longer timescales
would be required.
The bottom panel of Fig.\ \ref{sfr} shows the evolution of the SFR for
the stellar subgroups formed within SFRegG2 (cf.\ Sec.\ \ref{stars})
but clearly separate from G2 at almost all times. As is the case for
the whole G2 cluster, SF is quenched in most of the subgroups in the
feedback simulation LAF1. This is exemplified by the curve for
subG2, which indicates that SF is halted in this subgroup at $\tage
\approx 7$ Myr, which corresponds to $\sim 2$ Myr
after feedback starts to sweep the gas out from the central region
(see the solid magenta line in the left panel of Fig.\
\ref{fgasdense}). Thus, the quenching of SF is dominated by the feedback
from the most massive star, and, for more distant groups, quenching
begins once the expanding shell around that star has engulfed the region
containing each group.

\subsection{Cluster expansion}
\label{sec:clus_dispersal}

At the same time when SF is quenched in subG2 ($\tage \approx 7$ Myr),
the right panel of Fig. \ref{groups} shows that the stars in this
group (magenta dots) begin to expand.  This is therefore clearly an
effect of the removal of the gravitational potential of the gas from
the group, because the group has existed since $\sim 4$ Myr before, so
that most of its stars are already in place, and it does not expand
much at all in the non-feedback run LAF0. The same effect
occurs for all other secondary sites, and so, as can be seen from
the right panel of Fig.\ \ref{groups}, the combined expanding
motions of the main group and the subgroups produce a net expansion
of the cluster, but not as a simple radial expansion from the
central star-forming region, but rather as a {\it hierarchical expansion},
resulting from the combined expansion of all groups.

Additionally, the group indicated by the circles in Fig.\ \ref{fig1},
which is represented by the red dots in both panels of Fig.\
\ref{groups}, is seen to completely fall onto the center of mass by
$\tage \approx 10$ Myr in run LAF0, while its members remain at an
average distance of $\approx 2$ pc from the center of mass in run
LAF1. This is a pre-gas-removal effect, since in this case, the gas
clump is first pushed away from its infall path by the passing
expanding shell and then it forms stars. Therefore, in general, the
expansion of the cluster is the combined result of two effects: one is
the well known effect of the removal of the remaining gas in regions
that have already formed most stars, reducing the gravitational
potential after it has formed stars. In fact, the feedback may even
  cause accumulation of gas in the outskirts of
  the SF region, creating a new potential well that could
  pull the stars towards these new locations, a mechanism which
  \citet{ZA+19b} referred to as {\it gravitational feedback};
  however, we do not see an accumulation of dense gas
  in the periphery of the cluster in our simulation that
   could cause such an effect. The second is the indirect
injection of momentum to the stars by the feedback,
   which injects momentum to the gas from which the stars form,
    so that they inherit their motion from the gas. 

\subsection{Triggering versus inhibition}
\label{sec:trig_inhib}

An important question about the feedback is whether and to what extent
is trigger subsequent star formation episodes in nearby regions, as
originally proposed by \citet{Elmegreen+77}.  In relation to this
question, we focus on the formation of group subG6 (green dots in the
right panel of Fig.\ \ref{groups}) in run LAF1. This group has no
counterpart in the non-feedback run. This group forms at a distance
$\sim 9$ pc from G2 and at a time $\tage \sim 7.6$ Myr ($\sim 2.6$ Myr
after feedback started affecting gas and stars in the central parsec
of G2). An animation of the LAF1 run (not shown) shows that, as the
expanding shell collides with another infalling filamentary stream, a
locally gravitationally unstable clump forms, giving birth to subG2.
However, this is the only instance of stimulated star formation we
have found in the simulation with feedback, suggesting that this is
not the dominant mode of secondary star formation. Instead, all the
other subgroups form also in the non-feedback run, indicating that
their appearance was not triggered by the feedback from the main hub,
but rather was just the manifestation of the small-scale,
  low-mass peripheral collapsing regions in the GHC process. In this
sense, triggering in our simulation appears to be the exception rather
than the rule. This seems qualitatively
  opposite to what was reported
    by \citet{Dale+12}, who found
   that the negative effect of feedback may be compensated by
  the positive feedback in the form of triggered SF. We do not find
  such balance in our simulations. The reason
    for the disagreement is not clear at the moment, although it may
    be at least partly due to the very different morphology of our
    clouds and theirs, since ours are flattened due to their formation
  mechanism through the collision of streams, while theirs are not
  formed self-consistently, but initially set up as spherical
  clouds. In any case, a detailed comparison is in order, which we
  hope to carry out in a future contribution.

More recently, \citet{Bending+20} studied the role of
  photoionization feedback on large spiral arms environment,
    finding that the feedback causes the formation of larger masses of dense
    gas, and a period of increased star formation than in a simulation
    with no feedback, although the dense gas is more scattered, and
    ultimately they report a lower overall SFR. In our case, we do not
  see the formation of much additional dense gas, although this may be
an artifact of our simplified radiative transfer scheme. We plan to
perform detailed comparisons with simulations using full radiative
transfer to evaluate the long-term impact of our scheme.

\section{Discussion}
\label{disc}

\subsection{Implications}
\label{sec:implic}

The results presented in the previous section have a number of
implications for the study and understanding of the structure and
evolution of young stellar clusters. These can be summarized as
follows: 



First, our finding that most of the secondary star formation sites in
the periphery of the main hub are primordial rather than triggered
suggests that the vast body of star-forming sites that are proposed as
the result of triggering \citep[e.g.,] [see also the review by
Elmegreen 2011 and references therein] {Elmegreen+77, Peng+10,
  Samal+14, Ochsendorf+15, Duronea+17, Rugel+19, Panwar+19} may
actually contain a large fraction of cases where the star-formation
activity would have occurred anyway, with no need for triggering. The
GHC scenario implies that the star-forming sites in the clouds exhibit
a hierarchical nesting, in which a main star-forming hub is expected
to be surrounded by lower-mass star-forming sites \citep{VS+09,
  VS+19}. Since the main hub eventually produces massive stars, the
\hii\ regions they produce will eventually engulf the lower-mass
sites, and appear as if their star-forming activity may have been
triggered by the expanding shells.  Although we do not include
supernovae in our simulation, a similar effect is expected to occur
for cases where the trigger appears to be a supernova explosion
\citep[e.g.,] [] {Kounkel+18}. These authors note that proper motions
in the $\lambda$ Ori region imply a radial expansion, and that the
traceback age of this expansion exceeds the age of the youngest stars
near the outer edges of the region. They thus conclude that the
formation of those stars may have been triggered by a supernova
explosion when their parent regions were at half their present
distance from the cluster centre. According to our results, the
triggering is not indispensable, and the stars likely could have
formed at that location even without the passage of a shell.


In addition, according to our results, the most important effect of
the passage of an expanding shell through a secondary star-forming
site is the removal of most of its gas, with the corresponding
triggering of the expansion of the stars already formed there. Thus,
we propose that, for sub-groups undergoing {\it local} expansion, the
local traceback age of the sub-group may indicate the time of the
shell passage, rather than being an indication of a triggering event.


Finally, our result that the gas mass in the whole region increases
faster by accretion than it decreases by conversion to stars has the
unexpected implication that part of the observed smallness of the SFE,
as defined in eq.\ (\ref{eq:sfe}) is not entirely due to a low rate of
conversion to stars, but also to the fact that the star-forming
regions themselves are accreting mass from their environment, keeping
the ratio in the right-hand side of eq.\ (\ref{eq:sfe}) low.  This is
a crucial implication of the accretion at all scales in molecular
clouds expected to occur in the GHC scenario \citep{VS+19}.


\subsection{Comparison with previous work}
\label{other}

The evolution of clouds and their star formation activity in the GHC
scenario incorporates features of various models previously presented
in the literature. The fundamental premise of the GHC scenario
\citep{VS+19} is that the clouds are evolving in near pressure-less
collapse, because they contain a large number of Jeans masses.
Nevertheless, the clouds do contain moderately supersonic turbulent
motions which cause nonlinear density fluctuations. Therefore, the
collapse is extremely non-homologous, and regions of different density
evolve on their respective collapse timescales. These timescales are,
however, longer than the standard free fall time, $\tff = 3 \pi/(32 G
\rho)$, because strongly anisotropic objects collapse on timescales
longer than $\tff$ by factors that depend on the aspect ratio
\citep{Toala+12, Pon+12}. Since the collapse is nearly pressureless,
anisotropies are amplified, producing filamentary accretion flows that
feed the main star-forming hubs, although secondary, lower-mass star
formation occurs along the filaments as they become locally Jeans
unstable \citep{GV14}. These filaments are thus ``conveyor-belt'' type
of flows, as proposed by \citet{Longmore+14}. 

In addition, the primary
star formation activity in the hubs and the secondary activity  in the
filaments leads to a hierarchical process of star formation, in which
various groups are formed, thus being analogous to the hierarchical
cluster formation proposed by a number of groups \citep[e.g.,] []
{Fellhauer+09, Grudic+18b}. Our simulations are complementary to those
of the latter authors, because, on the one hand, we consider a
self-consistent formation and turbulent driving of the clouds and
stellar particles that represent individual stars, while they
considered initially spherical, rotating gas configurations, and their
sink particles represent stellar groups. On the other hand, those
authors considered a wide rang of cloud column densities, while we
consider only one case, corresponding to Solar-neighborhood-type
clouds. In particular, \citet{Grudic+18b} considered many cases of
densities so high that photoionising radiation alone is unable to
destroy the clouds, a case we do not encounter in our LAF1 simulation.

The GHC scenario represented by our simulations also implies an
acceleration of the star formation process, due to the global
gravitational contraction, with the corresponding increase of the mean
cloud density and decrease of the mean free-fall time \citep{ZA-VS14,
  VS+19}. The conclusion of an increasing SFR has also been reached by
\citet{Lee+15, Murray+15, Caldwell+18}, although \citet{Murray+15}
specifically suggest that the increase is due to local rather than
global collapse, arguing that there is no observational evidence for
global, cloud-scale collapse provided by P Cygni profiles toward GMCs
on scales larger than $\sim 10$ pc. As discussed in \citet{VS+19}, we
emphasize that, according to the GHC scenario, the infall does not
occur with a near-spherical geometry and thus should be not searched
by line-of-sight accretion tracers such as P Cygni profiles, but
rather through filamentary accretion flows, and these are routinely
observed now on scales up to 10 pc or more \citep[e.g.,] []
{Schneider+10, Kirk+13, Peretto+14, Tackenberg+14, Hacar+17, Chen+19}.
The larger-scale accretion flow from cloud to filaments, suggested by
the simulations \citep{GV14} is indeed more elusive, because it occurs
on lower-density gas for which the gravitational velocity is lower,
and thus more ``polluted'' by the turbulent background
\citep{Camacho+16}, but nevertheless a global velocity offset between
peripheral $^{12}$CO and internal $^{13}$CO has been detected for a
number of clouds by \citet{Barnes+18}.

Concerning the SFE, it has been measured in other
  numerical works for MCs of similar gas mass ($10^{4}$ \Msun) to that of
    our clouds, but with somewhat different results.
    For example \citet{Dale17} also finds SFEs that increase with time, but 
    reaching significantly larger values than we find here. 
    A possible reason, which we plan to
  explore in the future, is that his simulations start with {\it
    spherical} clouds, while our clouds acquire a sheetlike
  morphology, due to their formation by the collision of streams. This
implies that our clouds produce a much shallower potential well than
if their mass were assembled in a roughly spherical configurations.

It is also important to note that different
definitions for SFE are used in
  different works, and care must be taken to ensure that comparisons
  are meaningful. For instance, \citet{Geen+18}
 report ``total'' SFEs, with values in the range of 6-15
\%. To compute this total SFE, they take into account the total
initial mass in gas. Thus, to make a more adequate comparison, we can
use the maximum mass in dense gas attained by
SFReg2 throughout its evolution. In this case, the SFE in our simulation is
$\sim$ 9\% when measured 5 Myrs after feedback started to disperse the
gas, a similar value to that reported by \citet{Geen+18}.  

Our result of a quick destruction of the molecular cloud after the
first very massive star appears is consistent with the result by
\cite{Kim+18} that the destruction of a MC occurs within 2 --10 Myr
after the onset of radiation feedback. Along similar lines, 
\citet{Ali+19} reported that, for a MC with $10^{4}$ \Msun, 75 percent 
of its mass is dispersed within 4.3 Myr once a massive star of $\sim$ 34
 \Msun\ is placed in the most massive core within the cloud.
 In our case, this occurs after 5--10 Myr depending 
 of the star forming region within the numerical
box.  Similarly, \citet{Dale+12} found that photoionisation disperses
the neutral gas around 3 Myr before the explosion of the first
supernovae in clouds of mass $10^4$--$10^5$ \Msun.
\citet{Kim+18} also report that neutral gas is ejected at a typical
velocity of $\sim 6$-15 \kms. Although we do not track the velocity of
the dense gas directly, we do find (Sec.\ \ref{gas}) that the dense
gas is removed from the region at a velocity $\approx 11$ \kms\ from
the position of the massive star.
   
In relation to observations, it is interesting to cast our results in
the context, for example, of those by \citet{Ginsburg+16}. These
authors found that, in the W51 massive proto-custer clumps, the
feedback from massive stars may be insufficient to halt star formation
in the clumps, but may be capable of shutting off the large-scale
accretion onto them. Although our star-forming hub SFRegG2 is not as
massive, the images of Fig.\ \ref{fig1} suggest that indeed the
fueling of the hub is destroyed earlier than the hub itself (see the
second and third panels from the top on the right column), suggesting
a similar mechanism.

Our numerical results are also consistent with the timeline of GMC
evolution found observationally by \citet{Chevance+20} in the
PHANGS-ALMA survey (Leroy et al., in prep.), finding that GMCs
disperse within 1-5 Myr once massive stars emerge. As seen in Figs.\
\ref{fig1} and \ref{fgasdense}, the gas within 20 pc of SFRegG2 is
dispersed within $\sim 5$ Myr in our simulation. Moreover,
\citet{Chevance+20} also conclude that the total lifetime of GMCs is
10--30 Myr, and their final integrated star formation efficiency is
4--10\%. Our simulation results are fully consistent with these
numbers: the lifetime of the cloud is $\la 30$ Myr, as can be seen
from the fact that the first star forms at $t \sim 19$ Myr after the
beginning of the simulations, the first destructive massive star
appears $\sim 5$ Myr after the first star, and the cloud is destroyed
$\sim 5$ Myr after that, for a total of $\sim 30$ Myr of evolution.
This is an upper limit to the cloud's lifetime, since the cloud
becomes massive enough to be considered a GMC some $\sim 10$ Myr after
the start of the simulation. Thus, a better estimate of the lifetime
of our cloud is $\sim 20$ Myr, fully within the range found by
\citet{Chevance+20}. Also, the final SFE of our cloud is $\sim 5\%$,
also fully within the range of their findings. We conclude that our
simulation if fully consistent with their observational constraints.

\subsection{Caveats}
\label{caveats}

Although our simulations have the big advantage of forming stellar
particles with masses of individual stars thanks to our probabilistic
star formation prescription, the mass sampling of the IMF does not
reach the low-mass part of the IMF, since the minimum stellar mass we
have is $\sim 0.3$ \Msun, imposed by our numerical resolution. For a
typical IMF, about half of the total stellar mass is in stars with $M
< 3$ \Msun. It is not clear, however, whether our simulation is
missing that amount of mass, or has deposited it into the stellar mass
range it develops. In any case, the stars that do form have the
correct mass distribution, and thus the cluster $N$-body dynamics can
be trusted. Moreover, in regards to the effect of momentum injection to
the stellar component via the gas while it is forming stars, if the
effect is efficient for the higher-mass stars, it should be even more
so for the lower-mass ones. Finally, regarding the total amount of
energy injected, this does not depend on the lower-mass stars. Thus,
we conclude that our results should not be seriously affected by our
star-formation prescription, and is anyway more accurate than schemes
in which the sink particles have masses corresponding to small
clusters.

The other important caveat is the usage of our strongly simplified
prescription for injecting the feedback energy. Although in Paper I we
presented a test of the method in the standard problem of the
expansion of an \hii\ region in a uniform medium, showing that the
numerical solution was always within $\sim 30\%$ of the analytical
solution. In addition, the fact that our results are in qualitative
agreement with those of other groups, as discussed in Sec.\
\ref{other}, suggests that our implementation captures the essential
physics. Nevertheless, it is clear that our method has limitations,
and we intend to test our results in a forthcoming paper using a full
radiative-transfer simulation with the probabilistic star formation
prescription.

Finally, it is worth mentioning that we have analysed only one
  out of the three stellar clusters formed in the numerical box. It would certainly
  be desirable to have a large sample of simulated
  star-forming regions with different properties to perform
   a statistical analysis and test our conclusions.
  Unfortunately, the study
    presented in this paper could not be done in automated form, and
    had to be done manually, because it was necessary to keep track of
    each star's position even if it had migrated far from its
    birthplace (perhaps near another group, so that standard grouping
    algorithms could not be used), while simultaneously assigning the
    correct membership to the newly-formed stars, which required a
    constant updating of the definition of a group. A comprehensive
    analysis of a large sample of clusters will require first an
    automatization of the entire procedure.

\section{Summary and conclusions}
\label{conclusions}

In this paper, we have used two hydrodynamical simulations of MCs
undergoing global hierarchical collapse (GHC), with and without
feedback (LAF1 and LAF0, respectively), to investigate the effects of
the feedback on the shaping of the cluster, removal of the dense gas,
and the primordial or triggered nature of the peripheral star
formation activity. We focus on a region in the simulation labeled
SFRegG2, which contains approximately $10^{4}$ \Msun\ in dense gas,
and which forms stars in a hierarchy consisting of a main hub, denoted
G2, and several peripheral subgroups. A key ingredient of our
numerical simulations is the sub-grid model for star formation, which
allows the stellar particles to have realistic stellar masses, with a
Salpeter slope, in the range 0.39--61 \Msun, with each star feeding
back at the rate corresponding to its mass. This allows us to have a
realistic implementation of feedback from massive stars, and also a
realistic description of the cluster dynamics.

We found that the evolution of the two simulations is quite similar
until a very massive ($M \approx 30$ \Msun) forms, at $\tage \sim 5$
Myr after the formation of the first star in the cluster. After that
time, the expansion of the \hii\ region formed by this star dominates
the dynamics of both the gas and the stars in the region, causing the
feedback run to behave very differently from the non-feedback one. Our
main results are as follows:

\begin{itemize}

\item Because gas is funneled to the star-forming hubs and cores by the
  filaments, the amount of material available for forming the most
  massive stars is also limited, causing the most massive stars in run
  LAF1 to be less massive than the corresponding ones in run
  LAF0. Also, these most massive stars in run LAF0 appear at a later
  time than their less massive counterparts in LAF1, suggesting that
  more massive, denser clumps are necessary to form more massive
  stars.

\item Before the feedback from the very massive star begins to
  dominate, the mass accretion rate onto the star-forming region
  SFRegG2 is larger than the rate of gas consumption by star
  formation, so that the mass and density of the region increase, and
  the measured instantaneous SFE given by eq.\
  (\ref{eq:sfe}) remains low, in spite of the star formation activity
  occurring with virtually no opposition from feedback.

\item Most of the peripheral, low-mass star-forming sites form also in
  the non-feedback run LAF0, indicating they are of a primordial
  nature, rather than triggered by the passage of the expanding shell
  of the \hii\ region. We have found only one case of truly triggered
  star formation (i.e., a star forming site that occurs in the
  feedback run but not in the non-feedback one), out of a total of 6
  peripheral regions. This also implies that the net effect of the
  feedback on the SFR is to quench it, although locally it is possible
  to find examples of triggering.

\item After feedback begins to dominate, the dense gas is removed from
  the star-forming region SFRegG2 at a velocity $\approx 11\ \rm{km\
    seg}^{-1}$, as indicated by the times at which the dense gas mass
  contained within successively larger spheres around the central hub
  begins to decrease. The net flux of dense gas becomes
  outwards-directed, as shown in Fig.\ \ref{flux}, reversing the
  previous trend of the hub's mass increasing faster by accretion than
  it could consume its mass by star formation. During this stage, both
  the gas mass and the SFR in the region begin to decrease, although
  the onset of this decrease ``propagates'' outwards at the ejection
  velocity. Therefore, secondary star-forming sites in the periphery
  of the hub begin to lose their gas and reduce their SFR later than
  the central hub.

\item The spatial and kinematic structure of the stellar component is
  also strongly affected by the feedback, and not only because of the
  reduction of the local potential well caused by the removal of the
  dense gas. An important simultaneous mechanism is the injection of
  momentum to the star-forming gas in the peripheral sites by the
  expanding shell. In the non-feedback run, these sites are generally
  falling onto the main hub along the filaments, but the passage of
  the expanding shell partially counteracts this motion, reducing or
  reverting the infall. Therefore, the subgroups end up not having
  very large infall speeds towards the main hub, and instead have
  smaller, more random velocities.

\item The two previous results remove the concern about the GHC
  scenario raised by \citet[] [Sec.\ 3.4.1] {Krumholz+19}, that the
  stellar motions should be directed radially inwards before gas is
  expelled, and outwards after gas is removed. This criticisms arises
  from an oversimplification of the GHC scenario, ignoring its
  hierarchical nature and picturing it as a globally monolithic
  collapse. Our simulations show that the combination of the scattered
  nature of the hierarchical star formation region, with a main hub
  and several secondary neighboring sites, together with the
  ``braking'' of the infall by the pressure of the dominant \hii\
  region implies that: {\it a)} There is no strong global radial motion
  of the subgroups in relation to the central hub before the gas
  removal, and {\it b)} According to the discussion in Sec.\
  \ref{sec:clus_dispersal}, after each subgroup begins to lose its
  gas, the locally dominant expansion motion is with respect to its
  local center, not with respect to the main hub. The cluster as a
  whole expands as well, but as a consequence of the local expansion
  of each group, not as a coherent expansion from the central hub.
  Therefore, the resulting motions are much more moderate than
  estimated by \citet{Krumholz+19} under the assumption of a
  monolithic, rather than hierarchical, cloud collapse, and thus the
  resulting concerns are not applicable to GHC.

\end{itemize}

\section*{Acknowledgements}

A.G-S. was supported by DGAPA-UNAM Fellowship. This work has received
partial financial support from CONACYT grant 255295 to E.V.-S.

\section*{Data availability}

The data underlying this article will be shared on reasonable request to the corresponding author.


\bibliographystyle{mnras}

\bibliography{references}


%


\bsp	
\label{lastpage}
\end{document}